\documentclass[aps,prl,showpacs,floatfix,twocolumn,%
superscriptaddress]{revtex4-1}

\usepackage{amsmath}
\usepackage{amssymb}
\usepackage{graphicx}
\usepackage{bm}

\newcommand{\etapoly}{\eta_{\text{p}}}
\newcommand{\etasolv}{\eta_{\text{s}}}
\newcommand*{\br}{\ensuremath{\bm{r}}}

\newcommand{\Ninf}{\ensuremath{N\to\infty}}

\begin{document}

\title{Dynamic crossover scaling in polymer solutions}

\author{Aashish Jain}
\affiliation{Department of Chemical Engineering, Monash University,
Melbourne, VIC 3800, Australia}

\author{B. D\"unweg}
\affiliation{Max Planck Institute for Polymer Research,
Ackermannweg 10, 55128 Mainz, Germany}
\affiliation{Department of Chemical Engineering, Monash University,
Melbourne, VIC 3800, Australia}

\author{J. Ravi Prakash}
\email[Corresponding author: ]{ravi.jagadeeshan@monash.edu}
\homepage[Visit: ]{http://users.monash.edu.au/~rprakash/}
\affiliation{Department of Chemical Engineering, Monash University,
Melbourne, VIC 3800, Australia}

\date{\today}

\begin{abstract}
  The crossover region in the phase diagram of polymer solutions, in
  the regime above the overlap concentration, is explored by Brownian
  Dynamics simulations, to map out the universal crossover scaling
  functions for the gyration radius and the single-chain diffusion
  constant. Scaling considerations, our simulation results, and
  recently reported data on the polymer contribution to the viscosity
  obtained from rheological measurements on DNA systems, support the
  assumption that there are simple relations between these functions,
  such that they can be inferred from one another.
\end{abstract}


\maketitle

\begin{figure}
\begin{center}
\includegraphics[clip,keepaspectratio,width=0.47\textwidth]{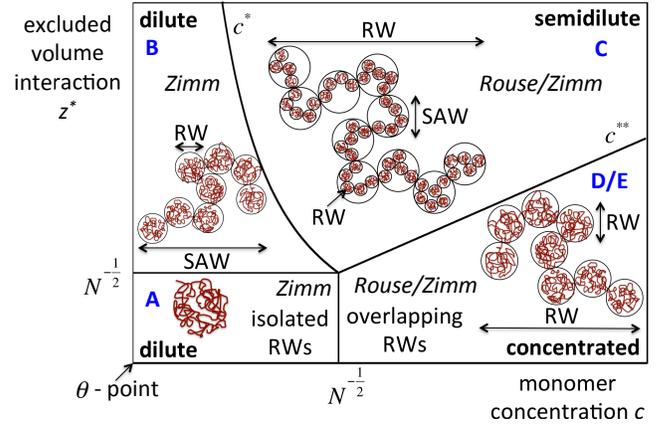}
\caption{(Color online) Phase diagram of a polymer solution in the
  plane monomer concentration $c$ and excluded-volume strength
  $z^\star$, as predicted by the standard blob picture. The $\Theta$
  regime occurs for small $z^\star \ll N^{-1/2}$, while $z^\star \gg
  N^{-1/2}$ in the good solvent regime.  The overlap concentration is
  given by $c^\star$. For concentrations above $c^{\star \star}$,
  excluded-volume interactions are completely screened.  For
  derivations of relations governing the crossover boundaries, see
  Supplemental Material~\cite{SupMat}.}
\label{fig:unscaled_phase_diagram}
\end{center}
\vskip-20pt
\end{figure}

The basic physics of systems of linear flexible uncharged
macromolecules in solution, in a wide range of concentrations and
solvent qualities, can be considered as reasonably well understood, at
least as far as the standard scaling laws are concerned, which govern
the static and near-equilibrium dynamic properties of the solution
\cite{dgen79,DoiEd86,GrosbergKhokhlov94,RubCol03}. The fairly rich
structure of the generic phase diagram of this system
\cite{GrosbergKhokhlov94,RubCol03}, shown in Fig.
\ref{fig:unscaled_phase_diagram}, is a result of the interplay between
molecule-molecule overlap on the one hand, and solvent quality effects
on the other. These factors determine whether excluded-volume (EV)
interactions between monomers are important or not. For highly dilute
systems, where overlap is unimportant, we may characterize the
effective strength of the EV interaction (or quality of the solvent)
by the dimensionless parameter $z^\star = ({\epsilon}/{k_B T})(1 -
\Theta / T)$, where $k_{B}$ is Boltzmann's constant, $T$ is the
temperature, $\Theta$ is the Theta temperature at which an isolated
infinitely long chain would collapse, and $\epsilon$ is the repulsive
energy (of enthalpic origin) with which two monomers interact. For
small $z^\star$, EV effects are unimportant, and the chain
conformations obey random-walk (RW) statistics ($R \sim b N^{1/2}$,
where $R$ is the size of the chain, $N$ its number of monomers and $b$
the monomer size), while for large $z^\star$ the chain is swollen and
obeys the statistics of a self-avoiding walk (SAW), $R \sim b N^\nu$,
where $\nu \approx 0.6$ is the Flory exponent. Upon increasing the
concentration $c$ (number of monomers per unit volume) beyond the
overlap value $c^\star$, Flory screening sets in, such that for such
systems again $R \propto N^{1/2}$, even in the case of a good
solvent. This interplay is well described by the ``blob'' concept
\cite{dgen79,GrosbergKhokhlov94,RubCol03}, where the thermal blob size
$\xi_T$ characterizes the length scale (in real space) up to which EV
is unimportant due to an insufficiently large value of $z^\star$,
while the overlap (or concentration) blob size $\xi_c$ is the typical
length scale above which chain overlap induces Flory screening. A
semidilute solution (regime C in
Fig. \ref{fig:unscaled_phase_diagram}) is characterized by $b \ll
\xi_T \ll \xi_c \ll R$, i.~e. by only a finite window of length scales
between $\xi_T$ and $\xi_c$ where SAW statistics apply. Upon further
increasing $c$, the system enters the concentrated regime $c \gg
c^{\star \star}$, where this window has shrunk to zero. The lines
drawn in Fig. \ref{fig:unscaled_phase_diagram} thus do not indicate
sharp phase transitions but rather smooth crossovers.

These crossovers in the static conformational properties are
accompanied by a dynamic
crossover~\cite{dgen79,DoiEd86,GrosbergKhokhlov94,RubCol03}. For $c
\ll c^\star$, hydrodynamic interactions (HI) are important, such that
the dynamics is characterized by the scaling laws of the Zimm model
(most notably $D \sim k_B T / (\etasolv R)$, where $D$ is the
single-chain diffusion constant, and $\etasolv$ the solvent
viscosity). For $c \gg c^\star$, HI are screened, essentially as a
result of chain-chain collisions which tend to randomize the momentum
propagation in the system. The fact that HI in semidilute solutions
are unscreened for short times~\cite{AhlDuenweg2001} (up to the Zimm
relaxation time of an overlap blob, which is the typical waiting time
for chain-chain collisions to occur) is of no relevance for the
present paper, since we are interested in the long-time dynamics only.
Therefore here the scaling laws of the Rouse model (e.~g. $D \propto
N^{-1}$) apply. We consider only systems where $N$ and $c$ are still
small enough that entanglements play no essential role, such that a
further crossover to reptation
dynamics~\cite{dgen79,DoiEd86,GrosbergKhokhlov94,RubCol03} does not
need to be taken into account. HI screening is governed by the same
length scale $\xi_c$ as EV
screening~\cite{deGennes1976,AhlDuenweg2001}.

Applying the blob picture to the various regimes of
Fig.~\ref{fig:unscaled_phase_diagram}, one finds straightforwardly the
scaling results for $\xi_T$, $\xi_c$, $R$, $D$, and the polymer
contribution to the viscosity $\etapoly$, listed in Table~I of the
Supplemental Material~\cite{SupMat}, and we refer the reader to the
literature~\cite{RubCol03} for further details.

\begingroup
\squeezetable
\begin{table*}
\begin{tabular}{| c | c | c | c | c | c |}
\hline
regime      & A
            & B
            & C
            & D
            & E
\\
\hline
\hline
$\xi_T / (b N^{1/2})$
            & --
            & $z^{-1}$
            & $z^{-1}$
            & --
            & --
\\
\hline
$\xi_c / (b N^{1/2})$
            & --
            & --
            & $\left( c / c^\star \right)
              ^{- \tfrac{\nu}{3 \nu - 1}} z^{2 \nu - 1}$
            & $\left( c / c^\star \right)^{-1} z^{3 (2 \nu - 1)}$
            & $\left( c / c^\star \right)^{-1}$
\\
\hline
$R / (b N^{1/2})$
            & $1$
            & $z^{2 \nu - 1}$
            & $\left( c / c^\star \right)^
              {- \tfrac{1}{2} \tfrac{2 \nu - 1}{3 \nu - 1}}
              z^{2 \nu - 1}$
            & $1$
            & $1$
\\
\hline
$D \etasolv b N^{1/2} / (k_B T)$         
            & $1$ 
            & $z^{- (2 \nu - 1)}$
            & $\left( c / c^\star \right)^
              {- \tfrac{1 - \nu}{3 \nu - 1}}
              z^{- (2 \nu - 1)}$
            & $\left( c / c^\star \right)^{-1} z^{3 (2 \nu - 1)}$
            & $\left( c / c^\star \right)^{-1}$
\\
\hline
$\etapoly / \etasolv$   
             & $c / c^\star$
             & $c / c^\star$
             & $\left( c / c^\star \right)^
               {\tfrac{1}{3 \nu - 1}}$
             & $\left( c / c^\star \right)^2
               z^{-6 (2 \nu - 1)}$
             & $\left( c / c^\star \right)^2$
\\
\hline
$R / R^\star$
             & $1$
             & $1$
             & $\left( c / c^\star \right)
               ^{- \tfrac{1}{2} \tfrac{2 \nu - 1}{3 \nu - 1}}$
             & $z^{- (2 \nu - 1)}$
             & $1$
\\
\hline
$D / D^\star$
             & $1$
             & $1$
             & $\left( c / c^\star \right)
               ^{- \tfrac{1 - \nu}{3 \nu - 1}}$
             & $\left( c / c^\star \right)^{-1}
               z^{4 (2 \nu - 1)}$
             & $\left( c / c^\star \right)^{-1}$
\\
\hline
$\etapoly / \etapoly^\star$
             & $c / c^\star$
             & $c / c^\star$
             & $\left( c / c^\star \right)^
               {\tfrac{1}{3 \nu - 1}}$
             & $\left( c / c^\star \right)^2
               z^{-6 (2 \nu - 1)}$
             & $\left( c / c^\star \right)^2$
\\
\hline
$\left( c / c^\star \right)^{1/4}
\left( R / R^\star \right)
\left( D / D^\star \right)^{1/4}$
             & $\left(c / c^\star \right)^{1/4}$
             & $\left(c / c^\star \right)^{1/4}$
             & $1$
             & $1$
             & $1$
\\
\hline
$\left( c / c^\star \right)^{1/3}
\left( R / R^\star \right)
\left( \etapoly / \etapoly^\star \right)^{-1/6}$
             & $\left( c / c^\star \right)^{1/6}$
             & $\left( c / c^\star \right)^{1/6}$
             & $1$
             & $1$
             & $1$
\\
\hline
\end{tabular}
\caption{
  Various normalized quantities of the polymer system (thermal blob
  size $\xi_T$, overlap blob size $\xi_c$, polymer radius $R$,
  single-chain diffusion constant $D$, polymer part of the
  viscosity $\etapoly$) in the regimes indicated in 
  Fig.~\ref{fig:scaled_phase_diagram},  
  in terms of  $c / c^\star$ and $z$. Blob sizes are not indicated 
  where they are irrelevant. Numerical prefactors of order unity have 
  been ignored. $R^\star$, $D^\star$, $\etapoly^\star$
  denote the values of $R$, $D$, $\etapoly$ at $c = c^\star$. For derivations, 
  see Supplemental Material~\cite{SupMat}.}
\label{tab:properties_scaled}
\end{table*}
\endgroup

In the past, these scaling laws (which are asymptotic results
applicable only sufficiently far away from the crossover regions), as
well as the concentration-driven crossovers, have been subjected to a
wealth of carefully executed experiments
\cite{RubCol03,wiltzius84,*EwenRichter97} and computer simulations
\cite{PaulBinder91,*MullerBinder2000,*gompper2010,AhlDuenweg2001}.
However, to the best of our knowledge, there have been only few
investigations that have tried to study the \emph{double} crossover
driven by both solvent quality and concentration, in particular when
dynamic properties are concerned~\cite{DaivisPinder90}. The present
Letter attempts to fill that gap. We are particularly interested in
the \emph{universal} crossover scaling functions~\cite{Schafer99},
which describe the behavior in the asymptotic limit $N \to \infty$.

\begin{figure}
\begin{center}
\includegraphics[clip,keepaspectratio,width=0.47\textwidth]{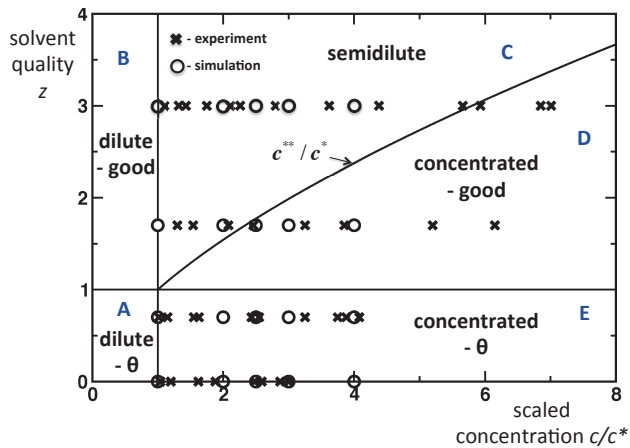}
\caption{(Color online) Same as Fig. \ref{fig:unscaled_phase_diagram}
  but with $c / c^\star$ and $z = z^\star N^{1/2}$ as variables. In
  this representation the loci of the crossovers do not depend on the
  chain length $N$. The points where the present simulations and
  recently reported experiments~\cite{pan2012} have been performed 
  are indicated as well.}
\label{fig:scaled_phase_diagram}
\end{center}
\end{figure}

The first step in analyzing universal crossover behavior is the
introduction of suitably scaled dimensionless variables (or ratios).
We therefore re-parameterize the scaling laws by replacing the EV
strength $z^\star$ with the ``solvent quality'' parameter $z = z^\star
N^{1/2}$ (note that this is the standard crossover scaling variable
used to describe the $\Theta$ transition), and the monomer
concentration $c$ with the ratio $c / c^\star$. The results of this
transformation are shown in Fig. \ref{fig:scaled_phase_diagram} and
listed in Table~\ref{tab:properties_scaled}. For $R$, $D$ and
$\etapoly$ we introduce dimensionless ratios as indicated in the
table. In this representation the dependence on the chain length $N$
has been completely absorbed in the $N$ dependence of $c^\star$ and
$z$, and the only remaining variables are $c / c^\star$ and $z$. It is
therefore natural to generalize these results to the relations $R /
R^\star = \phi_R ( c / c^\star, z )$, $D / D^\star = \phi_D ( c /
c^\star, z )$, and $\etapoly / \etapoly^\star = \phi_\eta ( c /
c^\star, z )$, where $R^\star$, $D^\star$, $\etapoly^\star$ denote the
values of $R$, $D$, $\etapoly$ at $c = c^\star$, and $\phi_D$,
$\phi_R$, and $\phi_\eta$ are universal crossover scaling functions
defined on the whole plane of Fig.  \ref{fig:scaled_phase_diagram},
and which, up to numerical prefactors, are known deep in the
asymptotic regimes (these are just the results listed in
Table~\ref{tab:properties_scaled}), but whose behavior needs to be
calculated or measured near the crossover lines.

An interesting observation can be made when looking at the last two
lines of Table~\ref{tab:properties_scaled}. In all regimes above the
overlap concentration, we find that the functions $R / R^\star$, $D /
D^\star$, and $\etapoly / \etapoly^\star$ are not independent, but can
be calculated as soon as one of them is known. It is then natural to
assume that the same property also holds in the crossover regimes,
i.e.\ that the relations $(c / c^\star)^{1/4} \phi_R \phi_D^{1/4} = \mbox{\rm
  const.}$ and $(c / c^\star)^{1/3} \phi_R \phi_\eta^{-1/6} = \mbox{\rm
  const.}$ hold in the whole plane $c / c^\star \gg 1$ --- or, in
other words, that there is essentially only a single crossover scaling
function in the problem. 

In what follows, we present results for $\phi_R$ and $\phi_D$ obtained
by careful computer simulations. It turns out that the convergence to
the universal behavior is quite rapid, and can be treated fairly
reliably by extrapolations
\cite{kroger2000,*Pra01,*KumPra03,*SunPra05,*SunPra06}. In addition,
we use recently reported data on $\phi_\eta$ (obtained from
rheological measurements on DNA systems)~\cite{pan2012}, to support
the hypothesis of the existence of a single crossover scaling
function.  Due to the universality of our results, they should be
broadly applicable to a large class of experimental systems.

A linear bead-spring chain model is used to represent polymer
molecules, with each polymer chain coarse-grained into a sequence of
$N$ beads connected by massless springs. An ensemble of $N_c$ such
bead-spring chains, enclosed in a cubic cell of edge length $L$ and
immersed in an incompressible Newtonian solvent, is used to model the
polymer solution. The monomer concentration of beads per cell is $c =
N_{\text{tot}} / V$, where $N_{\text{tot}} = N \times N_c$ is the
total number of beads, and $V = L^3$ is the volume of the simulation
cell. A Brownian dynamics algorithm~\cite{Jain2012} which accounts for
HI between the beads, and which can simulate both $\Theta$ and good
solvent conditions, is used here to obtain the stochastic time
evolution of bead positions. In particular, the pair-wise EV
interactions between beads $\nu$ and $\mu$ in a chain are represented
by a narrow Gaussian potential~\cite{praott99,KumPra03}
\begin{equation}
\frac{E \left( {\br}_{\nu \mu} \right)}{k_{\text{B}} T} = z^{\star}   \,
\left(   \frac{1}{ {d^{\star}}^3} \right)  \exp \left\lbrace - \frac{1 }{ 2 \,
b^2 }\, \frac{ \br_{\nu \mu}^2 }{ {d^{\star}}^2} \right\rbrace,
\label{evpot}
\end{equation}
where $d^{\star}$ is a non-dimensional parameter that measures the
range of excluded volume interaction, $b$ is the monomer size, and
${\br}_{\nu \mu}$ is the vector connecting monomers $\nu$ and $\mu$.
Simulations to explore the solvent quality crossover have been carried
out by keeping the value of $z$ constant at four different values: $\{
0, 0.7, 1.7, 3 \}$, with $z=0$ corresponding to $\Theta$-solutions. At
each fixed value of $z$, values of $c/c^\star$ ranging from 0.1 to 4
have been used to sample both the dilute and semidilute regimes. The
overlap concentration $c^\star$ is calculated from $c^\star = N /
\left[({4 \pi}/{3}) R_{g0}^3\right]$, where $R_{g0}$ is the radius of
gyration of a chain in the dilute limit, which is computed
\emph{apriori} by running \emph{single chain} BD simulations. The box
size $L$ is selected to ensure $L \geq 2 R_e$ (where $R_e$ is the
end-to-end distance of a chain), in order to prevent a chain from
wrapping over itself. For this purpose, the value of $R_e$ at any
value of $c/c^\star$ is estimated from the blob scaling law $R_e =
R_{e0} \left({c}/{c^\star}\right)^{- (2 \nu - 1)/(6 \nu - 2)}$ where
$R_{e0}$ is the end-to-end distance of a chain computed in the dilute
limit. Typical simulations consist of an equilibration run for
approximately one relaxation time followed by a production run of 10
to 60 relaxation times; here, the relaxation time $\tau$ was estimated
via $\tau = R_g^2 / (6 D)$, where $R_g$ is the concentration-dependent
radius of gyration. Moreover, data were obtained by averaging over
$30-70$ independent runs.

\begin{figure}
\begin{center}
\includegraphics[clip,keepaspectratio,width=0.47\textwidth]{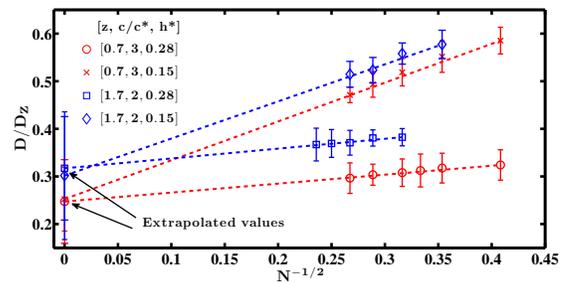}
\caption{(Color online) Universal predictions of the ratio $D/D_{Z}$
  at the two state points $(z, c/c^\star) = (0.7,3)$ and $(1.7,2)$,
  where $D$ is the single-chain diffusion constant at the state point
  under consideration, while $D_{Z}$ is the $c = 0$ value at the same
  $z$. Data accumulated for finite chain lengths $N$ (symbols) for two
  different values of the hydrodynamic interaction parameter
  ($h^{\star} = 0.15$ and $0.28$) extrapolate to a unique value in the
  limit $N \to \infty$. The limited range of $N$ values leads to a
  conservative estimate of the error of the extrapolated value.}
\label{fig:Extrapolation}
\end{center}
\end{figure}

Asymptotic results at any point $(c/c^\star,z)$, independent of
simulation parameters, are obtained by accumulating finite chain data
from BD simulations carried out at fixed values of these variables,
and subsequently extrapolating to the limit \Ninf. At finite values of
$N$, the results still depend on the parameters $d^\star$ [see
Eq.~(\ref{evpot})], and on the hydrodynamic interaction parameter
$h^\star = a / \sqrt{\pi (k_B T / H )}$, where $a$ is the bead radius
and $H$ is the spring constant. The parameter $d^\star$ is truly
irrelevant in the long-chain limit, i.e. asymptotic conformation
statistics are only governed by $c / c^\star$ and
$z$~\cite{kroger2000,*Pra01,*KumPra03,*SunPra05,*SunPra06}. Conversely,
$h^\star$ is a variable similar to $z^\star$, where the relevant
scaling variable governing the crossover from free-draining to
non-draining behavior is the draining parameter $h = h^\star
\sqrt{N}$~\cite{kroger2000,*Pra01,*KumPra03,*SunPra05,*SunPra06}. In
the present study we focus on the non-draining limit $h \to \infty$,
which is obtained by $N \to \infty$ at constant $h^\star$. This limit
is independent of $h^\star$, as demonstrated in
Fig.~\ref{fig:Extrapolation} for $D/D_{Z}$, which is the ratio of the
single-chain diffusion constant at a finite concentration to its value
in the dilute limit. Data for chains of lengths $N = 6$ to $N = 20$
have been extrapolated to $N \to \infty$.

\begin{figure}
\begin{center}
\includegraphics[clip,keepaspectratio,width=0.47\textwidth]{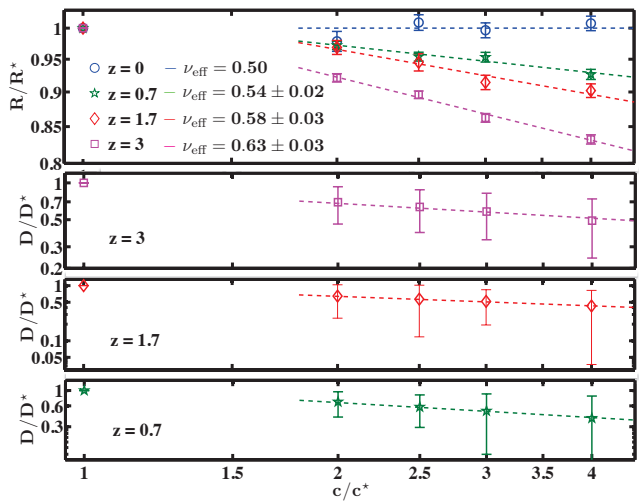}
\caption{(Color online) Concentration dependence of the crossover
  scaling functions $\phi_R$ and $\phi_D$ in the semidilute regime,
  for different values of solvent quality $z$, obtained by Brownian
  dynamics simulations.  The effective exponents $\nu_{\text{eff}}$
  have been determined by fitting power laws to the data $\phi_R
  \propto (c / c^\star)^{-p}$ and requiring $p = (2 \nu_{\text{eff}} -
  1) / (6 \nu_{\text{eff}} - 2)$, according to
  Table~\ref{tab:properties_scaled} in regime C. Using these values,
  lines are drawn according to $\phi_D \propto (c / c^\star)^{-q}$
  with $q = (1 - \nu_{\text{eff}}) / (3 \nu_{\text{eff}} -
  1)$. $\phi_D$ data for $z = 0$ are omitted because of large
  statistical errors.}
\label{fig:DbyDstarVscbycstar}
\end{center}
\end{figure}

\begin{figure}
\begin{center}
\includegraphics[clip,keepaspectratio,width=0.47\textwidth]{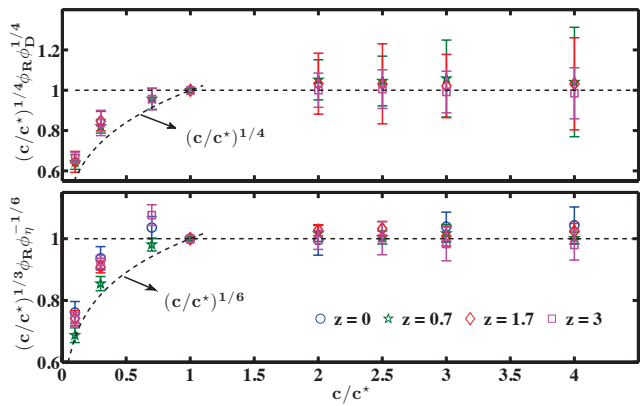}
\caption{(Color online) Demonstration that the functions $R / R^\star$, $D /
D^\star$, and $\etapoly / \etapoly^\star$ are not independent, but can
be calculated as soon as one of them is known. In the upper panel,
$z = 0$ data are omitted because of large statistical errors.}
\label{fig:etabyetastarScalingAll}
\end{center}
\end{figure}

The resulting crossover scaling functions $\phi_R( c / c^\star, z )$
and $\phi_D( c / c^\star, z )$ are shown in
Figs.~\ref{fig:DbyDstarVscbycstar}, where the mean polymer size $R$ is
$R_g$. The data are limited to the crossover region where in principle
none of the power laws of Table~\ref{tab:properties_scaled}
hold. Since the observation regime is small, it is nevertheless
possible to fit a power law to the data (without deeper theoretical
justification). This gives rise to a convenient description in terms
of an \emph{effective} empirical exponent $\nu_{\text{eff}}$ (for
values see Figs.~\ref{fig:DbyDstarVscbycstar}), which is obtained by
adjusting the value of $\nu$ in the scaling laws of regime C of
Table~\ref{tab:properties_scaled} to the observed slope. Since the
crossover scaling functions are related to each other (see below),
each crossover yields the same value for $\nu_{\text{eff}}$. Moreover,
consistency with the asymptotic laws requires that $\nu_{\text{eff}}$
varies between $0.5$ and $0.6$. A similar crossover scaling behavior
is obtained for $\phi_{\eta}( c / c^\star, z )$ from measurements of
$\etapoly$ for semidilute DNA solutions across a range of $N$, $c$,
and $z^\star$~\cite{pan2012}.

The validity of the expectation from scaling considerations that the
combination $(c / c^\star)^{1/4} \, \phi_R \, \phi_D^{1/4} = 1$ in the
entire semidilute regime, while it is proportional to $(c /
c^\star)^{1/4}$ in the dilute limit, and similarly that the
combination $(c / c^\star)^{1/3} \, \phi_R \, \phi_\eta^{-1/6} = 1$ in
the entire semidilute regime and $(c / c^\star)^{1/6}$ in the dilute
limit, is established in Figs.~\ref{fig:etabyetastarScalingAll}. In
the case of the latter combination, while values of $\phi_R$ have been
obtained from the present simulations, values of $\phi_\eta$ are from
the data reported in Ref.~\onlinecite{pan2012}. Clearly, the knowledge
of a single two-argument scaling function is sufficient to describe
the entire double crossover regime in polymer solutions.

\bibliography{crossover}

\end{document}


\title{Supplemental material for: \\
Dynamic crossover scaling in polymer solutions}

\author{Aashish Jain}
\affiliation{Department of Chemical Engineering, Monash University,
Melbourne, VIC 3800, Australia}

\author{B. D\"unweg}
\affiliation{Max Planck Institute for Polymer Research,
Ackermannweg 10, 55128 Mainz, Germany}
\affiliation{Department of Chemical Engineering, Monash University,
Melbourne, VIC 3800, Australia}

\author{J. Ravi Prakash}
\email[Corresponding author: ]{ravi.jagadeeshan@monash.edu}
\homepage[Visit: ]{http://users.monash.edu.au/~rprakash/}
\affiliation{Department of Chemical Engineering, Monash University,
Melbourne, VIC 3800, Australia}

\maketitle

\section{\label{scaling} Blob scaling in the concentration and 
solvent quality crossover regimes}

In regimes B and C of the phase diagram (see Fig.~1 in the main paper), where solvent quality effects play a role in determining the static and near-equilibrium dynamic properties of the solution, the length scale corresponding to thermal blobs is relevant. The size of the thermal blob $\xi_T$ can be estimated as follows.
Two monomers are assumed to interact with an energy $\epsilon \left( 1  - 
   \Theta/T   \right)$, $T > \Theta$, $\epsilon > 0$ --- i.~e. the
interaction is repulsive.  On small scales, the interaction is not
strong enough to perturb the random-walk (RW) nature of the chain. The chain notices that there is
repulsion only when the total repulsive
energy is of order $k_B T$. This defines the thermal blob size, which we assume corresponds 
to $n$ monomers. Since the number of contacts in a RW is of order
$n^{1/2}$ in three spatial dimensions, the condition to find $n$ is
\begin{equation}
\epsilon \left( 1  -    \frac{\Theta}{T} \right) n^{1/2} = k_B T .
\label{tblobcond}
\end{equation}
In the present work, the pair-wise excluded-volume (EV) interactions between monomers $\nu$ and $\mu$ in a chain are 
represented by a narrow Gaussian potential, 
\begin{equation}
\frac{E \left( {\br}_{\nu \mu} \right)}{k_{\text{B}} T} = z^{\star}   \,
\left(   \frac{1}{ {d^{\star}}^3} \right)  \exp \left\lbrace - \frac{1 }{ 2 \,
b^2 }\, \frac{ \br_{\nu \mu}^2 }{ {d^{\star}}^2} \right\rbrace,
\label{evpot}
\end{equation}
where, $d^{\star}$ is a non-dimensional parameter that measures the
range of excluded volume interaction, $b$ is the monomer size, and ${\br}_{\nu \mu}$ is the vector
connecting monomers $\nu$ and $\mu$. The parameter $z^{\star}$ is the non-dimensional strength of the pair-wise excluded-volume interactions. It follows that,
\begin{equation}
z^\star =  \frac{\epsilon}{k_B T} \left( 1  -     \frac{\Theta}{T} \right) .
\end{equation}
Equation~(\ref{tblobcond}) then implies,
\begin{equation}
n = (z^{\star})^{-2} .
\label{tblobmon}
\end{equation}
The corresponding blob size is consequently
\begin{equation}
\xi_T = b n^{1/2} = b  \, (z^{\star})^{-1} .
\label{tblob}
\end{equation}
Using the definition of the solvent quality parameter, 
\begin{equation}
z =z^{\star} N^{1/2},
\label{solqlty}
\end{equation}
where, $N$ is the number of monomers in a chain, the non-dimensional blob size in terms of 
\textit{scaled} variables is,
\begin{equation}
\frac{\xi_T}{ b N^{1/2}} = z^{-1}.
\label{tblob2}
\end{equation} 

In regimes B and C, on scales larger than $\xi_T$ excluded volume is important. In regime B, where 
the monomer concentration $c < c^\star$, the entire chain conformation is that of a self-avoiding walk (SAW) on length scales larger than  $\xi_T$. However, in regime C, where  $c > c^\star$, Flory screening sets in at yet larger length scales leading to the existence of concentration blobs. Thermal blobs within the concentration blob obey SAW statistics. If $\xi_c$ is the size of such a concentration blob, then its size is given by
\begin{equation}
\xi_c = \xi_T \, m^\nu ,
\label{cblob1}
\end{equation}
where, $\nu$ is the Flory exponent, while $m$ is the number of thermal
blobs within the concentration blob. The size $\xi_c$ is found
by the knowledge that the solution is homogeneous on scales larger
than $\xi_c$, and that the concentration blobs are space filling. The total number of monomers within $\xi_c$ is $n m$,
therefore 
\begin{equation}
n m  = {c} \, {\xi_c^3} .
\label{conc}
\end{equation}
Using Eqs.~(\ref{tblobmon}),  (\ref{cblob1}),  
and (\ref{tblob}), it is straight forward to show that 
\begin{equation}
m = \left( z^\star c^{-1} b^{-3} \right)^{1 / (3 \nu - 1)} .
\label{numblob}
\end{equation}
As a result, using Eqs.~(\ref{tblob}) and (\ref{numblob}) in Eq.~(\ref{cblob1})
\begin{equation}
\xi_c = b \, (z^{\star})^{- \tfrac{2 \nu - 1}{3 \nu - 1}} (c b^3)^{- \tfrac{\nu}{3 \nu - 1}}.
\label{cblob2}
\end{equation}
In order to represent the size of the concentration blob in terms of scaled variables, it is necessary to calculate the 
overlap concentration $c^\star$, which separates regimes B and C. The overlap concentration can be found 
from the condition that at $c = c^{\star}$, the concentration blob contains just all
monomers of the chain, $N$. From Eqs.~(\ref{conc}),  (\ref{cblob1}) and (\ref{tblob}), since $nm = N$ at $c = c^{\star}$
\begin{equation}
N = c^\star b^3 z^{\star - 3} m^{3 \nu} .
\end{equation}
Substituting for $m$ from Eq.~(\ref{numblob}), and solving for $c^\star$, leads to
\begin{equation}
c^\star =  b^{-3} N^{- (3 \nu - 1)} z^{\star - 3 (2 \nu - 1)} .
\label{cstar}
\end{equation}
When solved for $z^{\star}$, this leads to the equation of the line that 
separates regimes B and C in the unscaled variables phase diagram (Fig.~1 in the main paper). 

It is convenient to represent $c b^3$ as $(c/c^\star) (c^\star b^3)$ when converting from unscaled to scaled variables. One can then show from Eqs.~(\ref{cblob2}), (\ref{solqlty}) and (\ref{cstar}) that the non-dimensional size of the concentration blob in terms of scaled variables is
\begin{equation}
\frac{\xi_c}{ b N^{1/2}} =  \left( \frac{c}{c^\star} \right)^{- \nu / (3 \nu - 1)}  z^{2 \nu - 1}.
\label{cblob3}
\end{equation}

Upon systematically increasing the concentration $c$, the concentration blob shrinks until at a threshold concentration $c = c^{\star\star}$, $\xi_{c} = \xi_{T}$, and there is then no longer any length scale where SAW statistics apply. The locus of points $c^{\star\star}$ separates regime C from regimes D and E in the unscaled variables phase diagram. It is straight forward to calculate $c^{\star\star}$ by equating Eqs.~(\ref{tblob}) and (\ref{cblob2})
\begin{equation}
c^{\star\star} =  b^{-3} \,  z^{\star}.
\label{cdubstar1}
\end{equation}
In terms of scaled variables, it is similarly straight forward to show by equating Eqs.~(\ref{tblob2}) and (\ref{cblob3})
that
\begin{equation}
c^{\star\star} =   c^\star \, z^{2 (3 \nu - 1)} .
\label{cdubstar2}
\end{equation}

In regimes D and E, where only RW statistics apply, the concentration blob size is given by
\begin{equation}
{\xi_c} = { b g^{1/2}},
\label{cblobD}
\end{equation}
where, $g$ is the number of monomers in a concentration blob. Since the solution is homogeneous on all length scales larger than $\xi_{c}$, and the concentration blobs are space filling,
\begin{equation}
c  = \frac{g}{\xi_c^3} = \frac{g}{(b g^{1/2})^{3}}.
\label{DEconc}
\end{equation}
which implies 
\begin{equation}
g = (cb^{3})^{-2}
\label{Dblobmon}
\end{equation}
and as a result
\begin{equation}
\xi_c =  b \, (cb^{3})^{-1} .
\label{cblobD2}
\end{equation}
The expression for $c^{\star}$ given by Eq.~(\ref{cstar}) is only valid away from the $\Theta$ regime, \textit{i.e.}, for $z^{\star} > N^{-1/2}$. On the other hand, in the $\Theta$ regime, since at $c = c^{\star}$ the concentration blob contains all the monomers in the chain, which implies ${\xi_c} = { b N^{1/2}}$, Eq.~(\ref{cblobD2}) leads to 
\begin{equation}
c^{\star} =  b^{-3} \, N^{-1/2} .
\label{cstarE}
\end{equation}
This difference in the expressions for $c^{\star}$ makes it necessary to distinguish between regimes D (where EV effects are important) and E (where EV effects are negligible) when representing the size of the concentration blob (and indeed all other observables) in terms of scaled variables. In regime D, Eqs.~(\ref{cblobD2}), (\ref{cstar}) and (\ref{solqlty}) imply
\begin{equation}
\frac{\xi_c}{b N^{1/2}} =   \left( \frac{c}{c^\star} \right)^{-1} z^{3 (2 \nu - 1)} ,
\label{cblobD3}
\end{equation}
while in regime E, Eqs.~(\ref{cblobD2}) and (\ref{cstarE}) imply
\begin{equation}
\frac{\xi_c}{b N^{1/2}} =   \left( \frac{c}{c^\star} \right)^{-1}  .
\label{cblobE2}
\end{equation}

Once the blob scaling laws for $\xi_{T}$  are known in regimes B and C, and those for $\xi_{c}$ are known in regimes C to E, the scaling laws for all other observables in these regimes can be derived. Even though these laws have been derived and discussed in detail previously~\cite{dgen79,GrosbergKhokhlov94,RubCol03}, here, as examples, we derive the scaling laws for the static chain size $R$, and the single-chain diffusion coefficient $D$, in order to represent them in terms of the notation used in this work. The scaling laws for regime A are not discussed since the conventional notation in this regime is followed here. Once the expressions for $R$ and $D$ are known, the scaling laws for $\etapoly$, the polymer contribution to viscosity, can be derived from the relation $\etapoly \sim k_BT \, (c / N) \, \tau$, where $\tau$ is the longest relaxation time of the macromolecule, $\tau \sim R^2 / D$.

\begingroup
\begin{table*}[t]
\begin{tabular}{| c | c | c | c | c |}
\hline
regime      & A
            & B
            & C
            & D / E
\\
\hline
\hline
$\xi_T$     & --
            & $b \, (z^{\star})^{-1}$
            & $b \, (z^{\star})^{-1}$
            & --
\\ 
\hline
$\xi_c$     & --
            & --
            & $b \, (z^{\star})^{- \tfrac{2 \nu - 1}{3 \nu - 1}} 
              (c b^3)^{- \tfrac{\nu}{3 \nu - 1}}$
            & $b \, (c b^3)^{-1}$
\\ 
\hline
$R$         & $b N^{1/2}$
            & $b N^{\nu} (z^{\star})^{2 \nu - 1}$
            & $b N^{1/2} 
              \left[z^\star (c b^3)^{-1} \right]^
              {\tfrac{1}{2} \tfrac{2 \nu - 1}{3 \nu - 1}}$
            & $b N^{1/2}$
\\ 
\hline
$D$         & $\dfrac{k_B T}{\etasolv b N^{1/2}}$ 
            & $\dfrac{k_B T}{\etasolv b N^\nu (z^{\star})^{2 \nu - 1}}$
            & $\dfrac{k_B T}{\etasolv b N}
              (z^{\star})^{-2 \tfrac{2 \nu - 1}{3 \nu - 1}}
              (c b^3)^{- \tfrac{1 - \nu}{3 \nu - 1}}$
            & $\dfrac{k_B T}{\etasolv b N} (c b^3)^{-1}$
\\ 
\hline
$\etapoly$   & $\etasolv c b^3 N^{1/2}$
             & $\etasolv c b^3 N^{3 \nu - 1} (z^\star)^{3 (2 \nu - 1)}$
             & $\etasolv (c b^3)^{\tfrac{1}{3 \nu - 1}} N
               (z^\star)^{3 \tfrac{2 \nu - 1}{3 \nu - 1}}$
             & $\etasolv (c b^3)^2 N$
\\ 
\hline
\end{tabular}
\caption{Various quantities of the polymer system (thermal blob
  size $\xi_T$, overlap blob size $\xi_c$, polymer radius $R$,
  single-chain diffusion constant $D$, polymer part of the
  viscosity $\etapoly$) in the regimes indicated in Fig.~1 of the main paper, 
  as a function of monomer size $b$,
  chain length $N$, excluded-volume interaction strength $z^\star$,
  thermal energy $k_B T$, solvent viscosity $\etasolv$, and monomer
  concentration $c$, within the framework of scaling theory. Blob
  sizes are not indicated in cases where they are irrelevant.
  Numerical prefactors of order unity have been ignored. The
  scaling laws are valid in the asymptotic regimes sufficiently
  far away from the crossover boundaries.
}
\label{tab:properties_unscaled}
\end{table*}
\endgroup

\subsection{\label{B} Regime B}

Since the chain obeys SAW statistics on length scales larger than $\xi_{T}$,
\begin{equation}
R = \xi_{T} \, \left( \frac{N}{n} \right)^{\nu} .
\label{RB}
\end{equation}
Using Eqs.~(\ref{tblobmon}) and (\ref{tblob}), leads to
\begin{equation}
R = b N^{\nu} (z^{\star})^{2 \nu - 1} .
\label{RB1}
\end{equation}
The scaling relation in terms of scaled variables can be obtained by making use of Eq.~(\ref{solqlty})
\begin{equation}
\frac{R}{b N^{1/2}} = z^{2 \nu - 1} .
\label{RB2}
\end{equation}

Zimm dynamics are relevant on all length scales. From the Nernst-Einstein relation, $D ={k_B T}/ {\etasolv \, R}$, and Eq.~(\ref{RB1})
\begin{equation}
D = \frac{k_B T}{\etasolv \, b N^{\nu} (z^{\star})^{2 \nu - 1}},
\label{DB1}
\end{equation}
where, $\etasolv$ is the solvent viscosity. Defining a non-dimensional diffusion coefficient $D / (k_B T / \etasolv b N^{1/2} )$, it follows from Eqs.~(\ref{solqlty}) that it is given in terms of scaled variables by
\begin{equation}
\frac{D}{(k_B T / \etasolv b N^{1/2} )} = z^{-(2 \nu - 1)} .
\label{DB2}
\end{equation}

\subsection{\label{C} Regime C}

Since RW statistics are obeyed on length scales larger than $\xi_{c}$
\begin{equation}
R = \xi_{c} \, \left( \frac{N}{nm} \right)^{1/2} ,
\label{RC}
\end{equation}
where, $N/nm$ is the number of concentration blobs in a chain. Using Eqs.~(\ref{tblobmon}), (\ref{numblob}) and (\ref{cblob2}), leads to
\begin{equation}
R = b N^{1/2} \left[z^\star (c b^3)^{-1} \right]^{\tfrac{1}{2} \tfrac{2 \nu - 1}{3 \nu - 1}} .
\label{RC1}
\end{equation}
The scaling relation in terms of scaled variables is obtained by using Eqs.~(\ref{solqlty}) and~(\ref{cstar})
\begin{equation}
\frac{R}{b N^{1/2}} = \left( \frac{c}{c^\star} \right)^{- \tfrac{1}{2} \tfrac{2 \nu - 1}{3 \nu - 1}} z^{2 \nu - 1} .
\label{RC2}
\end{equation}

Rouse dynamics are obeyed on length scales larger than $\xi_{c}$. Therefore
\begin{equation}
D = \dfrac{k_B T}{\etasolv \, \xi_{c} \, (N/nm)} .
\label{DC}
\end{equation}
Using Eqs.~(\ref{tblobmon}), (\ref{numblob}) and (\ref{cblob2}), leads to
\begin{equation}
D = \dfrac{k_B T}{\etasolv b N} (z^{\star})^{-2 \tfrac{2 \nu - 1}{3 \nu - 1}} (c b^3)^{- \tfrac{1 - \nu}{3 \nu - 1}}.
\label{DC1}
\end{equation}
The non-dimensional diffusivity in terms of scaled variables follows from using Eqs.~(\ref{solqlty}) and~(\ref{cstar}) in Eq.~(\ref{DC1})
\begin{equation}
\frac{D}{(k_B T / \etasolv b N^{1/2} )} = \left( c / c^\star \right)^{- \tfrac{1 - \nu}{3 \nu - 1}} z^{- (2 \nu - 1)} .
\label{DC2}
\end{equation}

\subsection{\label{D} Regime D}

In regime D, RW statistics are obeyed on all length scales, both within the concentration blob, and by the concentration blobs themselves. As a result
\begin{equation}
R = \xi_{c} \left( \frac{N}{g} \right)^{1/2} = b N^{1/2}  ,
\label{RD}
\end{equation}
where Eqs.~(\ref{Dblobmon}) and~(\ref{cblobD2}) have been used for simplification. The non-dimensional chain size is trivially $R / (b N^{1/2}) = 1$. 

Rouse dynamics are obeyed on length scales larger than $\xi_{c}$. Using Eqs.~(\ref{DC}) (with $g=nm$), and Eqs.~(\ref{Dblobmon}) and~(\ref{cblobD2}) leads to
\begin{equation}
D = \dfrac{k_B T}{\etasolv b N} (c b^3)^{-1} .
\label{DD2}
\end{equation}
The non-dimensional diffusivity in terms of scaled variables follows from Eqs.~(\ref{solqlty}) and~(\ref{cstar})
\begin{equation}
\frac{D}{(k_B T / \etasolv b N^{1/2} )} = \left(\frac{ c}{  c^\star} \right)^{-1} z^{3 (2 \nu - 1)} .
\label{DD3}
\end{equation}

\subsection{\label{E} Regime E}

In regime E, the scaling relations for $R$ and $D$ in terms of unscaled variables are identical to those in regime D, since in both regimes, RW statistics are obeyed on all length scales, and Rouse dynamics apply on length scales larger than $\xi_{c}$. In terms of scaled variables, while the distinction between the two regimes is irrelevant for the static chain size, the difference in the expression for $c^{\star}$ due to the presence and absence of EV effects, respectively, manifests itself as a difference in the expression for the non-dimensional diffusivity. Using Eqs.~(\ref{solqlty}) and~(\ref{cstarE}), Eq.~(\ref{DD2}) reduces in terms of scaled variables to
\begin{equation}
\frac{D}{(k_B T / \etasolv b N^{1/2} )} = \left( \frac{ c}{  c^\star}  \right)^{-1} .
\label{DE}
\end{equation}

Table~\ref{tab:properties_unscaled} summarises all the unscaled equations derived in this section, including those for the polymer contribution to viscosity. Table~I in the main paper summarises the equations in terms of scaled variables.

\section{\label{diffusion} Estimation of the error in the diffusion coefficient $D$}

\subsection{\label{finite} Error at finite $N$}

The long time self-diffusion coefficient is calculated from the mean-square displacement (MSD) of the center of mass of each chain~\cite{Jain2012}. Each stochastic trajectory in the simulation leads to a time series for the MSD. An ensemble average over all the trajectories then gives the time series of the mean of the MSD (denoted here as $\text{MSD}_{\text{avg}}$). Since the initial $\text{MSD}_{\text{avg}}$ data represents transient short time diffusivity, while the data at large times have large error bars (they are based on a smaller number of blocks in the block averaging method), we typically discard the first $15-20\%$ and last $20-30\%$ of the data. This leads to a window of times $\Delta \tau$ in the $\text{MSD}_{\text{avg}}$ data that can be fitted with a straight line. At a given value of $N$, the size of $\Delta \tau$ depends on the state point $(z, c/c^\star)$. For a fixed set of parameters, the slopes of the lines fitted to each of the MSD trajectories in the ensemble, over the range of times $\Delta \tau$, is used here to obtain an ensemble of predicted diffusion coefficients. The mean diffusion coefficient $D$ and the standard error of mean is then determined from this ensemble.

\subsection{\label{extrap} Error in the extrapolated value at $N \to \infty$}

At each state point $(z, c/c^\star)$ the mean value of $D$ and the error in the mean, for a set of finite size chains with $N = 6, 8, \ldots, 20$,  is obtained as described above. The  mean and error-of-mean of the ratio $D/D_{Z}$ is then obtained via single chain BD simulations to determine $D_{Z}$ for the same values of $N$ and $z$. As described in the main paper, the asymptotic value of $D/D_{Z}$ in the long chain limit is obtained by plotting the data at finite $N$ as a function of $N^{-1/2}$, and extrapolating a straight line fit to the data, to the limit $N \to \infty$. The fitting of the straight line, and the estimation of the error in the extrapolated value is carried out here with the help of ``Least-Squares Fitting'' numerical routines provided in the GNU Scientific Library (GSL). 

Essentially, the GSL routine finds the least-squares fit by minimizing $\chi^2$, the weighted sum of squared residuals, for the straight line model $D/D_{Z} = c_0 + c_1 N^{-1/2}$. The weights are the inverse of the error at each data point. The fitting routine \textsf{gsl\_fit\_wlinear} returns the best-fit parameters $c_{0}$ and $c_{1}$, along with a $2 \times 2$ covariance matrix that measures the statistical errors on $c_{0}$ and $c_{1}$ resulting from the errors in the data. The standard deviations of the best-fit parameters are then given by the square root of the corresponding diagonal elements of the covariance matrix. Particularly conveniently, the routine \textsf{gsl\_fit\_linear\_est} uses the best-fit coefficients $c_0$, $c_1$ and their estimated covariance to compute the fitted function and its standard deviation at any desired point. By using \textsf{gsl\_fit\_linear\_est} to find the value of the fitted function and its error at $N^{-1/2}=0$, we determine both the  infinite chain length limit value of $D/D_{Z}$ and its associated error.

\section{\label{cpu} Scaling of computational cost with chain size}

As illustrated in Fig.~3 of the main paper, asymptotic predictions in the long chain limit have been obtained for each state point $ \left(z, { c}/{  c^\star} \right)$ by extrapolating finite chain data accumulated for chain lengths ranging from $N=6$ to $N=20$. The use of this rather limited range of $N$ values is necessitated by the computational cost of the Brownian dynamics simulation algorithm used here. As shown below, an estimate of the computational cost of the algorithm can be derived by using some simple scaling arguments, and the resulting expression can be verified by comparison with the CPU time required in the current simulations. 

The computational cost of carrying out an Euler integration of the governing stochastic differential equation \emph{for a single time step} has been shown in Ref.~\onlinecite{Jain2012} to scale with system size as  $N_{\text{tot}}^{x}$, where $x = 2.1$. Recall that $N_{\text{tot}} = N \times N_c$ is the total number of beads in a cubic cell of edge length $L$, with $N_c$ being the number of bead-spring chains. Since typical simulations consist of runs extending over several relaxation times $\tau$, followed by averaging over many independent runs, it is necessary to find the dependence of $\tau$ on $N$ in order to find the scaling of the total CPU cost. This is done as follows for simulations carried out in the semidilute regime C. 

From Eq.~(\ref{RC2}), the requirement that $L \ge 2 R$ in order to prevent chains from wrapping over themselves leads  to
\begin{equation}
L \sim  b  \left( \frac{c}{c^\star} \right)^{- \tfrac{1}{2} \tfrac{2 \nu - 1}{3 \nu - 1}} z^{2 \nu - 1} N^{1/2}.
\label{cellsize}
\end{equation}
The concentration of monomers in the simulation box is $c =  (N N_c) / L^{3}$. As a result, from Eq.~(\ref{cellsize})
\begin{equation}
c \sim  \left( \frac{N_{c}}{b^{3}} \right) \left( \frac{c}{c^\star} \right)^{ \tfrac{3}{2} \tfrac{2 \nu - 1}{3 \nu - 1}} z^{-3(2 \nu - 1)} N^{-1/2}.
\label{moncon}
\end{equation}
Using Eqs.~(\ref{cstar}) and (\ref{solqlty}), the overlap concentration $c^\star$ can be written in terms of the solvent quality $z$ as
\begin{equation}
c^\star \sim b^{-3} z^{- 3 (2 \nu - 1)} N^{-1/2}.
\label{cstscal}
\end{equation}
It follows from Eqs.~(\ref{moncon}) and (\ref{cstscal}) that $N_{c}$  is independent of $z$, and related to the scaled concentration through the relation
\begin{equation}
N_{c} \sim   \left( \frac{c}{c^\star} \right)^{ \tfrac{1}{2}  \tfrac{1}{3 \nu - 1}}.
\label{numchain}
\end{equation}
The number of chains in a simulation box is consequently \emph{constant} when $\left({c}/{c^\star} \right)$ is maintained constant.

The relaxation time of a macromolecule $\tau \sim R^2 / D$. From Eqs.~(\ref{RC2}) and (\ref{DC2}), this implies
\begin{equation}
\tau \sim \tau_{0} \left( c / c^\star \right)^{ \tfrac{2 - 3 \nu}{3 \nu - 1}} z^{3 (2 \nu - 1)} N^{{3}/{2}}.
\label{tau}
\end{equation}
where $\tau_{0} = \etasolv b^{3} / k_B T $ is the monomer relaxation time. 

\begin{figure}[t]
\begin{center}
\includegraphics[clip,keepaspectratio,width=0.47\textwidth]{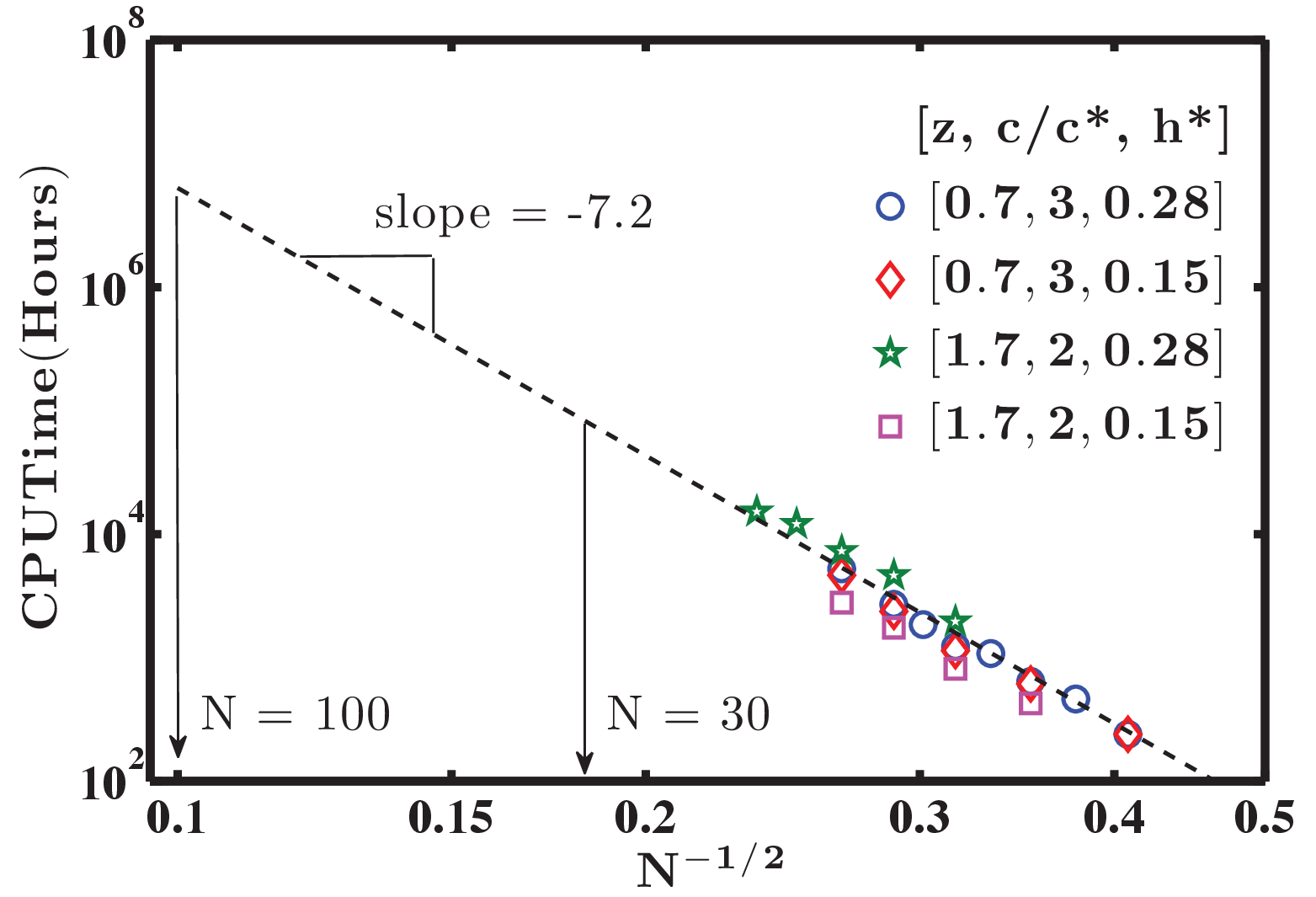}
\caption{(Color online) CPU time as a function of chain length for Brownian dynamics simulations carried out at the two state points $(z, c/c^\star) = (0.7,3)$ and $(1.7,2)$ for two different values of the hydrodynamic interaction parameter ($h^{\star} = 0.15$ and $0.28$), on a 156 SGI Altix XE 320 cluster. Symbols represent the various values of chain length $N$ used in the simulations, while the dashed line is drawn with a slope  predicted by the scaling relation Eq.~(\ref{totcpu}).}
\label{fig:cpucost}
\end{center}
\vskip-20pt
\end{figure}

The total computational cost of a single stochastic trajectory consequently scales as $\tau \, (N_{c} N)^{x}$. For fixed values of $z$ and $\left({ c}/{c^\star} \right)$ therefore
\begin{equation}
\text{Total CPU time per run} \sim N^{\tfrac{3}{2} + x} \sim \left(N^{-\tfrac{1}{2} }\right)^{-7.2}
\label{totcpu}
\end{equation}
where in the last expression on the right-hand-side we have substituted the value $x=2.1$ for the current algorithm, and used an exponent for $N$ that enables a representation of the CPU cost as displayed in Fig.~\ref{fig:cpucost}.

The various symbols in Fig.~\ref{fig:cpucost} represent simulations carried out at the two state points $(z, c/c^\star) = (0.7,3)$ and $(1.7,2)$, for two different values of the hydrodynamic interaction parameter $h^{\star}$, on a 156 SGI Altix XE 320 cluster. The simulations are identical to those used to display the ratio $D/D_{Z}$ in Fig.~3 of the main paper. It is immediately apparent that the prediction in Eq.~(\ref{totcpu}) for the  scaling of the total CPU time with chain size is obeyed closely by the present simulations, independent of solvent quality and scaled concentration. 

It is clearly desirable to add data for longer chains in order to improve the accuracy of the asymptotic value obtained by extrapolation to the long chain limit. Unfortunately, the additional CPU cost this would entail (which can be estimated from the scaling of computational cost with chain size shown in  Fig.~\ref{fig:cpucost}) makes this highly infeasible. For instance, simulating a chain with $N=30$ would require roughly $\mathcal{O}(10^{4})$ CPU hours, while $N=100$ would require roughly $\mathcal{O}(10^{6})$ CPU hours! Indeed, by including simulations for a chain with $N = 30$, the CPU time for obtaining the static and dynamic properties of a semidilute solution at a single state point $(z, c/c^\star) = (3,4)$ in the phase diagram (Fig.~2 of the main paper) is estimated to increase from its current value of approximately $ 7 \times 10^{4}$ hours to roughly $3 \times 10^{5}$ hours.

\begin{acknowledgments} 
  The authors gratefully acknowledge CPU time grants from the National
  Computational Infrastructure (NCI) facility hosted by Australian
  National University, and Victorian Life Sciences Computation
  Initiative (VLSCI) hosted by University of Melbourne. 
\end{acknowledgments}

\bibliography{crossover}